\documentclass[12pt,preprint]{aastex}
\usepackage[loose,ugly]{units}
\usepackage{myemulateapj5,apjfonts}
\RequirePackage{amssymb}
\RequirePackage{amsbsy}

\newcommand{\nuclei}[2]{\ensuremath{\mathrm{^{#1}#2}}}
\newcommand{\Hyd}{\ensuremath{\mathrm{H}}}

\newcommand{\C}{\nuclei{12}{C}}

\newcommand{\D}{\ensuremath{\mathcal{D}}}
\newcommand{\vdr}{\ensuremath{{w}}}

\newcommand{\Msun}{\ensuremath{M_\odot}}

\newcommand{\oneE}{\mbox{1E~1207.4-5209}}

\newcommand{\tgccm}{\ensuremath{\,\mathrm{\unit{g}\,\unit{cm^{-3}}}}}

\newcommand{\tgsqcm}{\ensuremath{\,\mathrm{\unit{g}\,\unit{cm^{-2}}}}}
\newcommand{\yrs}{\ensuremath{\,\unit{yrs}}}
\newcommand{\Kelvin}{\ensuremath{\,\unit{K}}}

\newcommand{\secs}{\ensuremath{\,\unit{s}}}
\newcommand{\CM}{\ensuremath{\,\unit{cm}}}
\newcommand{\Gauss}{\ensuremath{\,\unit{G}}}

\newcommand{\f}{\ensuremath{\mathcal{F}}}

\begin{document}

\title{Evolution of Young Neutron Star Envelopes}

\author{Philip Chang} 
\affil{Department of Physics, Broida Hall, University of California,
Santa Barbara, CA 93106; pchang@physics.ucsb.edu} 
\and 
\author{Lars Bildsten} 
\affil{Kavli Institute for Theoretical Physics and Department of
Physics, Kohn Hall, University of California, Santa Barbara, CA 93106;
bildsten@kitp.ucsb.edu}

\begin{abstract}

  We extend our initial study of diffusive nuclear burning (DNB) for
neutron stars (NSs) with Hydrogen atmospheres and an underlying layer
of proton capturing nuclei. Our initial study showed that DNB can
alter the photospheric abundance of Hydrogen on surprisingly short
timescales ($10^{2-4}\yrs$). Significant composition evolution impacts
the radiated thermal spectrum from the NS as well as its overall
cooling rate. In this paper, we consider the case when the rate
limiting step for the H consumption is diffusion to the burning layer,
rather than the local nuclear timescale.  This is relevant for NSs
with surface temperatures in excess of $10^6\,{\rm K}$, such as young
($<10^5$ yr) radio pulsars and accreting NSs in quiescence.  When
downward diffusion is the limiting rate in DNB, the rate of H
consumption is suppressed by 1-2 orders of magnitude compared to a DNB
estimate that assumes diffusive equilibrium. In order to apply our
ongoing study to young neutron stars, we also include the important
effects of strong magnetic fields ($B \sim 10^{12}\,{\rm G}$). In this
initial study of magnetic modifications to DNB, we find that the H
burning time is lengthened by 2-3 orders of magnitude for a
$10^{12}\,{\rm G}$ field. However, even for NSs with dipole field
strengths of $10^{12}$ G, we find that all of the H can be burned
before the pulsar reaches an age of $\sim 10^5 \ {\rm yr}$, thus
potentially revealing the underlying proton-capturing
elements. Finally, we conclude by providing an overview of what can be
learned about fallback and pulsar winds from measuring the surface
composition of a young NS.

\end{abstract}


\section{Introduction}

 The surface compositions of young NSs are now regularly probed by
Chandra and XMM-Newton observations (see Pavlov, Zavlin \& Sanwal
2002a for a review; also see Sanwal et al. 2002; Mereghetti et
al. 2002; Bignami et al. 2003). Since the amount of matter in the
photosphere is small ($\sim 10^{-20} \Msun$), there is no reliable way
to predict the composition of these outer layers from supernova
theory. Fall-back during the supernovae event (or later) could provide
a large range of possible elements (Woosley \& Weaver 1995), ranging
from pure H (due to spallation of heavier elements; Bildsten, Salpeter
\& Wasserman 1992), mid-weight material (due to a ``soft-landing'' of
outer shells falling back), or iron (if no fallback occurred).  

The current X-ray observations of young radio pulsars suggest that
those younger than $\sim 10^{4-5}\,{\rm yrs}$ possess magnetic H or He
atmospheres (Pavlov, Zavlin \& Sanwal 2002a), whereas pulsars older
than $\sim 10^{4-5}\,{\rm yrs}$ possess heavier element atmospheres
(see Pavlov et al. 2002b and references therein; also see Kaminker,
Yakovlev \& Gnedin 2002). This suggests a possible evolution of H or
He to more blackbody-like elements on a timescale of $10^{4-5}\,{\rm
yrs}$, and motivated our work (Chang \& Bildsten 2003; hereafter CB03)
on a mechanism of nuclear evolution called diffusive nuclear burning
(DNB) (first mentioned by Chiu \& Salpeter 1964 and later calculated
by Rosen 1968) that provides for a surprisingly rapid depletion of H.

The physics of DNB is very simple. At the photosphere, the local
temperature ($T_e \sim 10^6\,{\rm K}$) and density ($\rho \sim 1
\,{\rm g}\,{\rm cm}^{-3}$) are far too low to drive nuclear burning.
However, 1 meter underneath the photosphere, the density and
temperature are much higher ($\rho \sim 10^{5-6}\,{\rm g\,cm^{-3}}$,
$T \sim 10^8\,{\rm K}$) so any surface H that diffuses down to this
region will be rapidly captured onto a heavier nucleus.  For example,
for H capture onto Carbon, the $^{13}{\rm C}$ produced in this
reaction (following the $\beta$ decay of $^{13}$N) sinks into the
NS. This H depletion at depth then drives a H current from the surface
that, over time, can completely deplete the H at the photosphere. The
surprising realization is that the consumption of H is set by burning
which occurs in the exponentially suppressed diffusive tail, and can
easily consume all of the H on a NS in timescales of $\sim 10^5$
years.

In CB03, we found that at the characteristic depth, $y_{\rm burn}$
(i.e. the burning layer), where the H is rapidly consumed, there was
adequate time for diffusive equilibrium to be established for $T_c < 5
\times 10^7\,{\rm K}$ for H on C.  Hence at these low $T_c$'s (and
therefore low $T_e$'s), the rate limiting step is nuclear processes
rather than diffusion. However, at higher $T_c$'s typical of young
NSs, the nuclear timescale, $\tau_{\rm nuc}$, rapidly drops below the
diffusion timescale, $\tau_{\rm diff}$, so that diffusive equilibrium
no longer holds and downward diffusion to the burning layer is the
rate limiting step. The simple estimate for this transition to
diffusion-limited DNB is when $\tau_{\rm nuc} \sim \tau_{\rm diff}$,
where $\tau_{\rm diff} = h^2/\D$ is the time for H to diffuse an ion
scale height, $h$.  We find that incorporating the effects of
diffusion into DNB decreases H consumption by 1-2 orders or magnitude
or greater at high $T_c$ compared to a naive calculation assuming
diffusive equilibrium.  In addition, the dependence of the rate of H
consumption with $T_c$ changes when diffusion is the rate limiting
step.  Because of the size of the effects of diffusion at high $T_c$
on DNB, we only use an approximate treatment of diffusion in our
estimates of DNB rates.

In addition, we also include the modification of the thermal profile
due to the opacity changes induced by such high $B$ fields (Ventura \&
Potekhin 2001; Potekhin \& Yakovlev 2001), since nearly all young
neutron stars are also strongly magnetic ($B \sim 10^{12}\,{\rm
G}$).  We find that strong magnetic fields alter the burning rate due
to the modification of the thermal profile by 2-3 orders of magnitude.

   Applying our estimates to young NSs, we find that the age after
which DNB is no longer active is roughly $10^{5-6}$ years for standard
cooling (and much sooner for rapid cooling).  Prior to the cessession
of DNB, H is always depleted on a timescale shorter than the present
age.  Following Michel (1975), we show that pulsar wind excavation of
the NS surface is potentially a powerful mechanism for composition
evolution.  In particular, the pulsar wind can excavate to depths that
exceed the maximum thickness allowed for a H or He layer due to
thermonuclear constraints.

This paper is structured as follows.  In \S~2 we discuss the basic
equations and their modification in the limit where diffusion is the
rate limiting step. We discuss the effects of the magnetic field on
the envelope's thermal structure and DNB in \S~3, which then allows us
to apply our work to young neutron stars in \S~4.  We also speculate
on the possible excavation of material of NSs due to ion loss in a
pulsar wind.  We conclude in \S~5 by summarizing our results and
providing our views on what can be learned from measurements of young
NS surface composition.

\section{Diffusion Limited DNB}\label{DNB:diffusion_limited}

We only consider the top $10^4$ \CM\ of the envelope, where the
density is $< 10^{10} \tgccm$. The thickness of this layer is $\ll R$,
so we assume a plane parallel atmosphere with constant downward
gravitational acceleration, $g$. Consider an envelope consisting of
protons, electrons and ions that are coupled to each other and obey
charge neutrality ($n_e = \sum_i Z_i n_i$).  The equations of
hydrostatic balance are,
\begin{mathletters}\label{eq:general_hb}
\begin{eqnarray}
\frac {dP_p} {dr} &=& -n_p \left[ m_p g - e E + r_{ep}
n_e (v_p - v_e) + r_{ip} n_i Z_i^2 (v_p - v_i)\right], \\
\frac {dP_i} {dr} &=& -n_i \left[ A_i m_p g - Z_i e E + r_{ep} n_e
  Z_i^2(v_i - v_e) + \right.\nonumber \\
& & \left. r_{ip} n_p Z_i^2 (v_i - v_p)\right], \\ 
\frac {dP_e} {dr} &=& -n_e\left[m_e g + e E + r_{ep}
n_i Z_i^2(v_e - v_i) + r_{ep} n_p (v_e - v_p)\right],
\end{eqnarray}
\end{mathletters}
where $P_i$, $n_i$, $A_i$, $Z_i$, $v_i\equiv J_i/n_i$ are the
pressure, number density, atomic number, charge and relative velocity
of the $i$'th ion species in Eulerian coordinates and $E$ is the
upward pointing electric field.  General relativistic effects have
been ignored.  For simplicity, we presume the ions obey an ideal gas
equation of state, neglecting effects such as plasma nonideality.  We
also ignore thermal diffusion effects, which are expected to play only
a minor role in determining the structure of the NS envelope as the
temperature scale height is larger than the ion scale height. In
addition the thermal diffusion coefficient is small and hence thermal
diffusion effects are expected to be minor compared to ordinary
diffusion or gravitational settling in dense plasmas (Paquette et
al. 1986 in the case of trace fully-ionized C in fully-ionized He).
Plasma non-ideality effects are more significant especially for
determining the structure of an envelope with two or more species with
the same A/Z (i.e. He, C, N, O) and is an interesting question that
needs to be addressed in the future.  In the dense regions of the
plasma, plasma non-ideality effects are expected to come in as $P_{\rm
Coulomb}/P$, where $P_{\rm Coulomb}$ would be the Coulomb pressure and
$P$ is the local pressure, which could be large (as large as 20\%) in
the region of interest. Plasma non-ideality effects could be
particularly significant in the case of highly magnetized plasmas
($B>10^{11}\,{\rm G}$) as the Fermi pressure is decreased by the
presence of a quantizing magnetic field (Lai 2001).  Hence our model
is very simplified, but captures the essential physics of this
problem.  All vectorial quantities are positive radially. The coupling
coefficient, $r_{jl}$, is due to scattering of species j by species l.
In CB03, we consider the case where these velocities are small, so
that these drag terms could be neglected.  However, this is not the
case here.

We consider the case where the protons are a drifting trace and the
ions and electrons are a fixed background so $v_e \approx v_i \approx
0$ and $n_e \approx Z_i n_i \gg n_p$.  Equations (\ref{eq:general_hb})
simplify to
\begin{mathletters}
\begin{eqnarray}\label{eq:hb} 
\frac {dP_p} {dr} &=& -n_p \left( m_p g - e E + r_{ip} n_i Z_i^2 v_p\right), \\
\frac {dP_i} {dr} &=& -n_i \left( A_i m_p g - Z_i e E\right), \\ 
\frac {dP_e} {dr} &=& -n_e\left(m_e g + e E\right).
\end{eqnarray}
\end{mathletters}
We have dropped the drag terms in the ion and electron hydrostatic
balance equations since the drag force exerted on them by the protons
is $\ll m_p g$. The drag term in the proton hydrostatic balance
equation is only relevant in the presence of a nontrivial proton
current, $J_p \equiv n_p v_p$.  The drag coefficient is simply related
to the diffusion coefficient, $\D$, by taking the definition of the
current,
\begin{equation}\label{eq:def_current}
J_p = -\D \frac {dn_p} {dr} + n_p \vdr_p = n_p v_p, 
\end{equation}
where $\vdr_p$ is the proton drift velocity, and relating it to our
equations of hydrostatic balance. To begin let us presume that no
current exists ($J_p = 0$) so that equation (\ref{eq:def_current})
becomes $d\ln n_p/dr = \vdr_p/\D$.  For an isothermal atmosphere the
proton hydrostatic balance equation is
\begin{equation}
\frac {d\ln n_p} {dr} = \frac {-m_p g + eE}{k_B T},  
\end{equation}
which when equated to that from the diffusion equation gives
$\vdr_p/\D = (-m_p g + eE)/k_B T$. For $J_p \ne 0$, the proton
equation of hydrostatic balance is
\begin{equation}\label{eq:proton_hb}
\frac {d\ln n_p} {dr} = \frac {-m_p g + eE - r_{ip} n_i Z_i^2 v_p}{k_B T},
\end{equation} 
whereas the diffusion equation becomes
\begin{equation}\label{eq:proton_diff}
\frac {d\ln n_p} {dr} = \frac {\vdr_p - v_p} {\D}.
\end{equation}
Taking our earlier expression for $\vdr_p$ and equating equations
(\ref{eq:proton_hb}) and (\ref{eq:proton_diff}) yields the relation, 
\begin{equation}\label{eq:rFromD}
r_{ip} = \frac {k_B T} {n_i Z_i^2\D},
\end{equation}
between the drag and diffusion coefficients.

In the limit of ideal gases, when Coulomb corrections to the ion
equation of state are small ($\Gamma \equiv Z^2 e^2/(a k_B T) < 1$,
where $a$ is the interionic spacing), we parameterized \D\ using the
results of Chapman \& Cowling (1952) (also see Alcock \& Illarionov
1980). In the liquid regime ($1 < \Gamma < 175$), where most of the
burning occurs, we begin with the fit of Hansen, McDonald \& Pollock
(1975) of the dimensionless self diffusion coefficient, $\D^* =
\D_s/(\omega_p a^2)2.95 \Gamma^{-4/3}$, where $\omega_p$ is the plasma
frequency and $a$ is the mean ion separation.  We compute the
interspecies diffusion coefficient from the self-diffusion coefficient
in the same manner as Brown, Bildsten \& Chang (2002).  We assume the
Stokes-Einstein relation, $\D_s/k_B T = (4\pi a_1 \eta)^{-1}$, where
$a_1$ is the charge neutral sphere of the background species and
$\eta$ is the viscosity.  The diffusion coefficient for a trace in a
background is $\D/k_B T = (4\pi a_2 \eta)^{-1}$, where $a_2$ is the
charge neutral sphere around the trace.  Hence we find $\D = \D_s
a_1/a_2$, which gives
\begin{equation}\label{eq:diff_coeff}
  \D \approx 10^{-3} \frac{A_1^{0.1} T_6^{1.3}}{Z_1^{1.3}
    Z_2^{0.3} \rho_5^{0.6}} \,\CM^2\,\sec^{-1},
\end{equation}
where 1 and 2 are the background and trace ion respectively, $T_6 =
T/10^6\Kelvin$ and $\rho_5 = \rho/10^5 \tgccm$. The burning layer is
usually in the liquid region unless the effective temperature is
extremely high or low.  The self-diffusion coefficient fitted by
Hansen et al. (1975) is off by 50\% at $\Gamma\approx 1$, compared to
the more complete calculation of Paquette et al. (1986), but is
accurate to within a few percent for $\Gamma>10$.  The work of
Paquette et al. (1986) presents a more consistent way of bridging the
transition from the dilute approximate of Chapman \& Cowling and the
strong coupling results of Hansen et al. (1975), but for our estimates
equation (\ref{eq:diff_coeff}) is adequate in the liquid regime. At
extremely low $T_e$'s, the burning layer crystallizes ($\Gamma > 175$
for a one-component plasma; Potekhin \& Chabrier 2000) and halts
diffusion.
 
The thermal structure is determined by the heat diffusion equation for
a constant flux,
\begin{equation}
\frac {dT} {dr} = -\frac {3 \kappa \rho} {16 T^3}T_e^4,
\label{eq:flux}
\end{equation}
where $\kappa$ is the opacity.  The radiative opacity is determined
mostly by free-free absorption, where the Gaunt factor, $g_{\rm ff}$,
is fitted by Schatz et al. (1999).  We use the conductive opacities
given by Baiko et al. (1998) with analytic formulae given by Potekhin
et al.  (1999a).  For NSs with low surface magnetic fields ($B < 10^9
\Gauss$), we use the Paczynski (1983) electron equation of state
(which is accurate to within 5\% for ideal electron plasmas).

The continuity equation for the protons is
\begin{equation}
\frac {\partial n_p} {\partial t} + \frac {dJ_p} {dr} = -\frac {n_p}
{\tau_{\rm p, nuc}},
\label{eq:continuity}
\end{equation}
where $\tau_{\rm p, nuc}$ is the local proton lifetime against nuclear
capture.  As we have argued in CB03, the timescale for changing the
local number density is the timescale for changing the Hydrogen
column, which is much longer than the local nuclear timescale.  Hence,
we take a steady state approximation ($\partial n_p/\partial t = 0$),
which provides the current induced by DNB.  We combine equations
(\ref{eq:def_current}), (\ref{eq:proton_hb}) and (\ref{eq:continuity})
to obtain
\begin{equation}\label{eq:from_hb}
\frac {d^2 n_p} {dr^2} + \frac{dn_p} {dr} l_p^{-1} - n_p\frac {r_{ip}
  n_i Z_i^2} {k_B T \tau_{\rm p, nuc}} = 0,
\end{equation}
where $l_p = k_B T/(eE-m_p g)$ is the proton scale height.

Another way to derive equation (\ref{eq:from_hb}) is to combine the
proton continuity equation (eq.  [\ref{eq:continuity}]) with the
definition of the current (eq.[\ref{eq:def_current}]) to obtain
\begin{equation}\label{eq:full_diff}
-\frac {d^2n_p} {dr^2} + \frac {\vdr_p} {\D} \frac {dn_p} {dr} + \frac
{n_p} {\D \tau_{\rm p, nuc}} = 0,
\end{equation}
where for the purposes of this discussion we presume \D\ and $\vdr_p$
are constants.  Our full analysis includes the position dependence of
\D\ and $\vdr_p$ in equation (\ref{eq:full_diff}). However, we will
argue that only two terms really matter in the final analysis.  Taken
with equation (\ref{eq:rFromD}), we see that equation
(\ref{eq:full_diff}) is equivalent to equation (\ref{eq:from_hb}).

Care must be taken when solving equation (\ref{eq:full_diff}) since
$\tau_{\rm p, nuc}$ has a powerful temperature dependence.  We write
$n_p = \f_p\,n_{p,0}$ in terms of a scaling factor $\f_p$ and
$n_{p,0}$, the solution in diffusive equilibrium.  After dividing by
$n_{p,0}$, equation (\ref{eq:full_diff}) becomes,
\begin{eqnarray}
&\frac {\f_p}{n_{p,0}}\left(-\D\frac{d^2 n_{p,0}} {d r^2} +
    \frac {d n_{p,0}} {d r}\vdr_p\right) + \frac{d 
\f_p}{d r}\left(-\D\frac{d \ln n_{p,0}}{d r} + \vdr_p\right) -\nonumber\\
&\D\frac{d^2
\f_p}{d r^2} - \D\frac{d 
\f_p}{d r}\frac{d \ln n_{p,0}}{d r} = -\frac {\f_p} {\tau_{\rm p, nuc}}.
\end{eqnarray}
The first term in parentheses is the homogeneous diffusion equation
and hence it is zero, while the second term, $-{d\ln n_{p,0}}/{d r} +
l_p^{-1}$, is the ion equation of hydrostatic balance and is solved by
$n_{p,0}$, making this term zero as well.  Thus the full diffusion
equation reduces to
\begin{equation}\label{eq:ansatz_reduced}
\frac{d^2 \f_p}{d r^2} + \frac{d \f_p}{d r}\frac{d \ln n_{p,0}}{d r} =
\frac {\f_p} {\D \tau_{\rm p, nuc}},
\end{equation}
which we solve with the boundary conditions $\f_p \approx 1$ and $d
\f_p/d r \approx 0$ far above the burning layer.  Since $\tau_{\rm p,
nuc}$ is a nuclear timescale, which falls very rapidly as we move into
the deeper, hotter regions of the envelope, the nuclear scale height,
$l_{\rm nuc}^2 \equiv \D \tau_{\rm p, nuc}$, rapidly decreases into
the envelope.  Hence the second term in equation
(\ref{eq:ansatz_reduced}) rapidly becomes insignificant as the proton
scale height, $l_p = \left(d\ln n_{p,0}/dr\right)^{-1}$ becomes $ \gg
l_{\rm nuc}$ at the burning layer.  Thus for all practical purposes,
the second term could be dropped and we would have a diffusion
equation for $\f_p$ with a nuclear driven source.  Therefore, $\f_p$
defines the departure from diffusive equilibrium at a given depth as a
result of nuclear burning. If we include the position dependence of
\D\ and $\vdr_p$, this would not change our result significantly since
inclusion of those terms would introduce additional terms with $l_p$
and the thermal scale height, $l_T = \left(d\ln T/dr\right)^{-1}$,
both of which are $\gg l_{\rm nuc}$ at the burning layer.  The total
diffusive current is given by
\begin{equation}\label{eq:col_burning_rate}
\zeta_H = \frac {y_H} {\tau_{\rm col}} = \int \frac {\f_p n_{p,0}} {
  \tau_{\rm p, nuc}} dr = \int \frac {\f_p n_{p,0}} {
  \tau_{\rm p, nuc}} \frac y {\rho} d\ln y, 
\end{equation}
where $y = P/g$ is the total column density, $\rho$ is the local
density and $\tau_{\rm col}$ is the H column lifetime.

\begin{figure}
\epsscale{1.0}
    \plotone{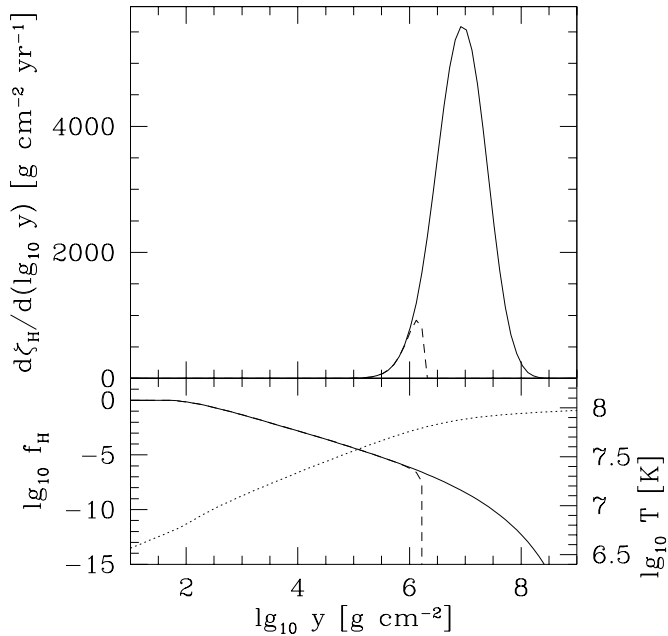}
    \caption{Total burning rate of H taking into account p-p capture
      and p + \C\ capture.  The bottom panel shows the number
      fraction, $f_H = n_H/n_{\rm tot}$, and temperature (dotted line)
      as a function of column.  This particular model has a total H
      column of $100 \tgsqcm$ with an effective temperature of $1.5
      \times 10^6 \Kelvin$. For the burning rate and number fraction,
      the dashed lines are the solutions taking diffusion limiting
      into account.  The solid lines are the solutions assuming
      diffusive equilibrium (CB03). Assuming diffusive equilibrium,
      the burning rate of this model is $\zeta_H \approx 6300
      \tgsqcm\,{\rm yr}^{-1}$. When diffusion limiting is taken into
      account, the burning rate is $\zeta_H \approx 400 \tgsqcm\,{\rm
      yr}^{-1}$.}
    \label{fig:col_vs_burning}
\end{figure}

Figure \ref{fig:col_vs_burning} shows the thermal and compositional
structure of the envelope (eq. [\ref{eq:hb}], [\ref{eq:flux}] and
[\ref{eq:ansatz_reduced}]) and the total burning rate, $\zeta_H$.  We
also solve the model assuming full diffusive equilibrium (CB03).  The
burning cuts off the H number concentration at large columns.  The
exponential increase in burning rates modifies the solution for the
total diffusive current by a factor of two compared to the case of
full diffusive equilibrium.  We showed that the condition of DNB in
diffusive equilibrium is satisfied when the temperature in the burning
layer $T(\rho) < 4.55 \times 10^7 \rho_5^{-0.11}\,{\rm K}$ (CB03).
The particular model that we have chosen ($T_e = 1.5 \times 10^6
\Kelvin$) is in the diffusion limited regime and illustrates the
reduction in the burning rate when diffusion becomes the rate limiting
step.  Compared to the calculation assuming diffusive equilibrium,
incorporating diffusion effects decreases the burning rate, $\zeta_H$,
by a factor of 25.  This difference between a naive calculation which
assumes diffusive equilibrium with one that includes the effects of
diffusion gets even larger at high $T_c$'s.

\begin{figure}
\epsscale{1.0} 
  \plotone{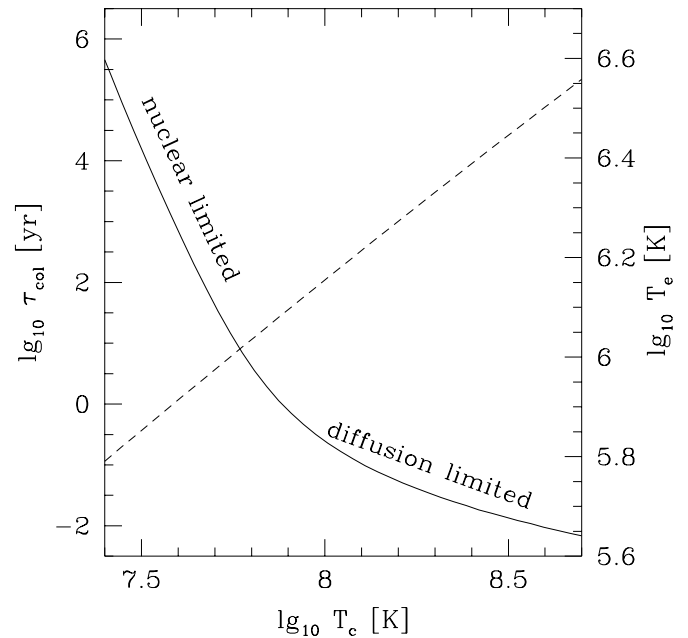}
  \caption{Lifetime (solid line) of Hydrogen with a fixed total
    column, $y=100\tgsqcm$ on Carbon and effective temperature (dashed
    line) as a function of core temperature.  The model with effective
    temperature, $T_e = 10^6 \Kelvin$, or core temperature, $T_c = 5.5
    \times 10^7 \Kelvin$, represents the transition from the nuclear
    limited to diffusion limited regimes of DNB for H on C.}
  \label{fig:Te_vs_tau_diff}
\end{figure}

The effect of incorporating the effects of diffusion into the DNB
calculation is illustrated by Figure \ref{fig:Te_vs_tau_diff}, which
shows the H column lifetime, $\tau_{\rm col}$, as a function of core
temperature, $T_c$.  In nuclear limited DNB, the column burning rate
follows the scalings derived in CB03 for \Hyd/C, $\tau_{\rm col}
\varpropto y_H^{-5/12} T_c^{-9.8}$ (for a range of $T_c$ between
$2-4\times 10^7\,{\rm K}$, but as noted in CB03, the scaling varies
depending on the temperature we expand about; for lower temperatures,
$T_c < 2 \times 10^7\,{\rm K}$, the exponential scaling with $T_c$ is
-12 to -13).  As we transition to diffusion limited DNB, the scaling
with central core temperature becomes more gentle as shown in Figure
\ref{fig:Te_vs_tau_diff}. The scaling of $\tau_{\rm col}$ with total H
column, $y_H$, remains the same as in nuclear limited DNB, since the H
number density in the burning layer scales like $\varpropto
y_H^{-\delta_i}$ (CB03), where $\delta_i = (Z_i + 1)/A_i - 2 \approx
-1.42$ for H on C, and is totally independent of the details of the
burning.

\section{Effects of Strong Magnetic Fields on DNB}

In order to apply our work to radio pulsars, we must include the
effects of strong magnetic fields ($B \sim 10^{12}\ {\rm G}$) on
the microphysics of the NS envelope.  Magnetic fields of $>10^{11}\
{\rm G}$ affect the thermal structure of the envelope by modifying the
electron equation of state and opacity (for a review see Lai 2001 and
references therein).  The changes to the electron equation of state
modify the electric field, which modifies the diffusion problem.
However, we neglect magnetic effects on the ions since magnetic fields
$> 10^{16}\,{\rm G}$ would be required to modify the ion EOS.

To model the full range of the electron EOS and opacity, we adopt the
formalism of Potekhin \& Yakovlev (2001) and references therein to
relate the electron pressure and electron number density to the
chemical potential.  We calculate the chemical potential from the
local pressure by modifying the iterative procedure described by
Potekhin \& Yakovlev (1996).  The free-free opacity in a magnetic
field is from equation (21) and (22) of Potekhin \& Yakovlev (2001).
For electron conduction opacities, we use the results of Potekhin
(1999). Starting from the photosphere at $\tau=2/3$, our numerical
integration of the constant flux equation (\ref{eq:flux}) for the
magnetized envelope agree to within 1\% with the thermal profiles
(i.e. $T(\rho)$) given by Potekhin et. al. (2003) for accreted
magnetized envelopes.

\begin{figure}
\epsscale{1.0}
  \plotone{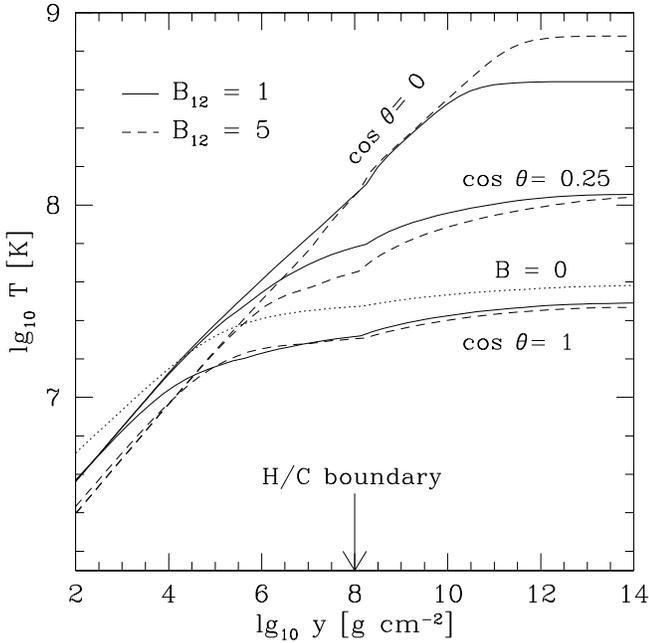}
  \caption{Thermal structure of a magnetic neutron star envelope for
    various values of field strength and inclination where $\cos
    \theta = 1$ is the magnetic pole. The effective temperature is
    fixed at $T_e = 10^6\ {\rm K}$ and a H/C envelope is assumed with
    the H/C boundary at $y = 10^8\ {\rm g\ cm^{-2}}$.  The thermal
    structure for a B=0 envelope is shown by the dotted line.}
    \label{fig:col_vs_T_Bfield}
\end{figure}

The thermal structure of a Hydrogen on Carbon magnetized envelope is
given in Figure \ref{fig:col_vs_T_Bfield} for magnetic fields of
various strengths and orientation, which we define in terms of
$\theta$, the angle of the magnetic field relative to the radial
direction.  These NSs have surface gravities of $g = 2.43\times
10^{14} \CM\secs^{-2}$, for a 1.4 \Msun, 10 km NS with an effective
temperatures of $T_e = 10^6 \Kelvin$.  We also show the nonmagnetic
case (dotted line) for purposes of comparison. Compared to the
nonmagnetic case, the magnetic envelopes are different in several
respects.  For a magnetic field with a significant radial component (
i.e. $\cos \theta = 1$ or $0.5$), the base temperature is lower for a
given effective temperature compared to the nonmagnetic case, while
for more horizontally oriented magnetic fields, the base temperature
is higher for a given effective temperature, highlighting the effect
of magnetic field orientation on the opacity of the envelope.  The
transition from primarily radiative to primarily conductive heat
tranport, after which the envelope becomes isothermal, takes place at
larger column and hence higher density compared to the nonmagnetic
case.  For the magnetic cases, orientation rather than field strength
is a stronger determininant of the thermal profile, because the
opacities along the field lines are a 4-7 orders of magnitude less
than opacities across the field lines (for a review see Potekhin \&
Yakovlev 2001 and references therein).  Therefore the temperature
distribution is highly latitude dependent, with the star hotter at the
magnetic poles and cooler at the magnetic equator.

The electric field determines the Hydrogen concentration, $f_H$, in
the Carbon layer (CB03) and is a function of the adiabatic index,
$\chi_{\rho} = (d\ln P_e/d\ln \rho) \varpropto (d\ln P_e/d\ln n_e)$.
In the absence of magnetic effects, the adiabatic index undergoes a
smooth transition from $\chi_{\rho} = 1$ in the ideal gas case to
$\chi_{\rho} = 5/3$ in nonrelativistic degeneracy to $\chi_{\rho} =
4/3$ in relativistic degeneracy.  However, in the presence of a
quantizing magnetic field, the adiabatic index also undergoes de
Haas-van Alphen oscillations in the weakly quantizing regime
(Potekhin, Chabrier \& Shibanov 1999b; Lai 2001) that also appears in
the electric field.

\begin{figure}
\epsscale{1.0}
  \plotone{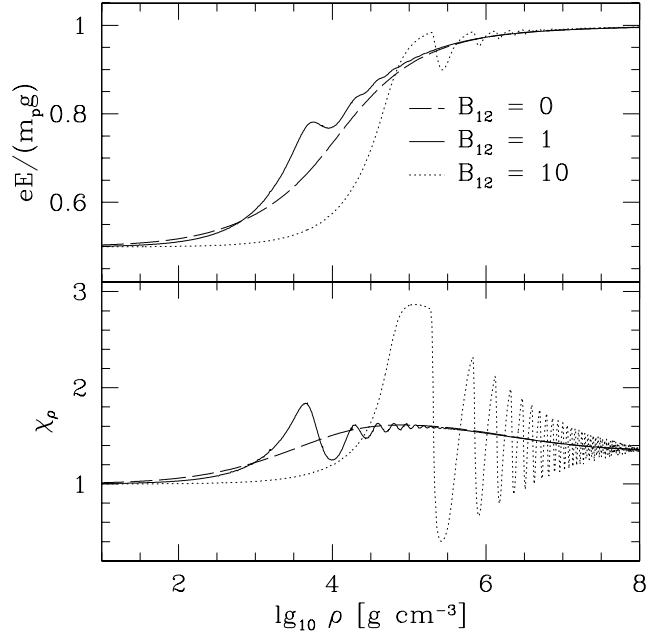}
  \caption{Electric field and adiabatic index $\chi_{\rho} = d\ln
    P_e/d\ln \rho$ for a pure H envelope for various B-fields.
    The shift of the degeneracy point for the higher B-field case is
    readily apparent for the $B = 10^{13} \Gauss$ case.  The initial
    approach of the adiabatic index to three for the higher B-field
    case and the response of the de Haas-van Alphen oscillations to
    B-field strength is also shown.  As expected all three cases
    asymptotically approach the relativistic degenerate limit of
    $eE/m_pg = 1$ and $d\ln P_e/d\ln \rho = 4/3$.}
    \label{fig:EfieldDiag}
\end{figure}

Figure \ref{fig:EfieldDiag} shows the adiabatic index, $\chi_{\rho}$,
and electric field as a function of density for the nonmagnetic case
($B=0$), $B=10^{12}\,{\rm G}$ and $B = 10^{13}\,{\rm G}$.  The de
Haas-van Alphen oscillations are evident as the electron EOS adiabatic
index climbs from the value for the ideal gas equation of state to
that of a 1-D Fermi gas, $P_e \varpropto \rho^3$.  As the Fermi
energy, $E_F$ exceeds the integer values of the cyclotron energy, the
equation of state softens, leading to alternating rising and falling
of the adiabatic index and electric field.  These oscillations are
smeared out at the larger densities and temperatures deeper in the NS
as $E_F$ greatly exceeds the cyclotron energy and we recover the
nonmagnetic solution.  

\begin{figure}
\epsscale{1.0}
  \plotone{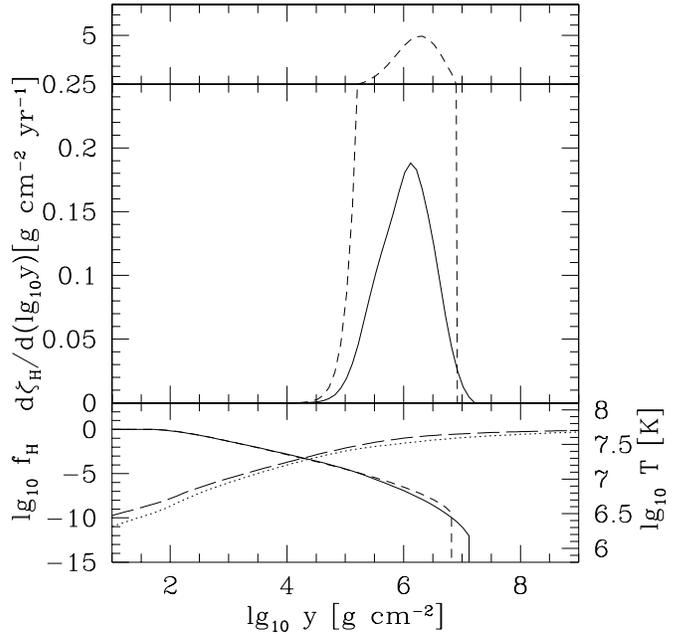}
  \caption{Total burning rate of H taking into account p-p capture and
    p + \C\ capture for a magnetized envelope with $B=10^{12}\,{\rm
    G}$ at the magnetic pole ($\cos\theta = 1$) and unmagnetized
    envelope with the same core temperature, $T_c = 5.5\times
    10^7\,\Kelvin$, and the same total H column, $y_H = 100\,\tgsqcm$.
    The bottom graph shows the number fraction and temperature as a
    function of column for the magnetized (solid and dotted lines) and
    unmagnetized (dashed and long-dashed lines) envelopes.  The
    nonmagnetic envelope has a column lifetime of $\tau_H \approx 7
    \,{\rm yr}$, while the magnetic envelope has $\tau_H \approx
    200\,{\rm yr}$ at $B=10^{12}\,{\rm G}$.}
  \label{fig:col_vs_burning_bfield}
\end{figure}

For simplicity, we have ignore the effects of a strong magnetic field
on diffusion. However, we expect the effects of magnetic fields on
diffusion in the burning region is of little importance at typical
pulsar fields.  The following estimate illustrates this point. The
mean free path for proton diffusion across magnetic fields is roughly
given by the proton-gyro radius which is given by $r_{g,p} =
\sqrt{E/m_p}/\omega_B \approx 10^{-8} (T/10^8\,{\rm K})^{1/2}
B_{12}^{-1}\,{\rm cm}$, where $\omega_B = eB/(m_pc)$ is the cyclotron
frequency.  The mean free path for proton diffusion is given by the
mean ion spacing $a =$$ [3\rho/$ $(4\pi A m_p)]^{-1/3} \approx 4 \times
10^{-10} \rho_5^{-1/3} A^{1/3}\,{\rm cm}$.  Since the ion spacing is
typically an order of magnitude smaller than the proton-gyro radius,
we expect the effects of typical pulsar magnetic field on diffusion in
the burning region to be small. Magnetic effects on diffusion is
significant in the low density regions of the envelope ($\rho \sim
10\,{\rm g\,cm}^{-3}$), but changes to the diffusion coefficient in
this region is of little importance to diffusion-limited DNB. For
magnetar strengths, this is no longer the case and the effects of
magnetic fields on diffusion must be taken into account.

In Figure \ref{fig:col_vs_burning_bfield}, we plot the burning
integrand, $d\zeta_H/(d\ln y)$, for a model envelope with $T_e = 1.125
\times 10^6 \Kelvin$, $B = 10^{12} \Gauss$ at the magnetic poles
($\cos\theta = 1$) and compare this against unmagnetized model with
the same core temperature and same total H column, $y_H$.  Compared to
the unmagnetized model, the magnetized model at $B=10^{12}\,{\rm G}$
burns at a rate that is an order of magnitude less compared to the
unmagnetized model. The reason for this lowered burning rate is a
combination of lowered H number density and lowered temperature at the
burning layer.  The bottom plot of Figure
\ref{fig:col_vs_burning_bfield} illustrates both points.  At the
burning layer (which is roughly the same for the magnetized and
unmagnetized cases), the magnetized case for $B=10^{12}\,{\rm G}$ has
both lowered temperature and lowered H number density.  Another
crucial difference is that, despite possessing the same core
temperature, $T_c = 5.5 \times 10^7 \Kelvin$, the magnetic envelope
($B=10^{12}\,{\rm G}$) has a higher effective temperature ($T_e
\approx 1.1 \times 10^6\,{\rm K}$) than the nonmagnetic case ($T_e
\approx 10^6 \Kelvin$) due to the lower radiative opacity of highly
magnetized plasmas.

Because the H number density or concentration and temperature profile
are both modified by the presence of a quantizing B-field, the
magnetic effects on DNB are nontrivial.  The B-field always lowers the
temperature at a \emph{fixed column} for fixed $T_c$ as shown in the
bottom plot of Figure \ref{fig:col_vs_burning_bfield}.  The higher
effective temperature of the magnetized envelope is wholly due to the
increase in depth where the photosphere sits. If the burning layer was
at a fixed column, this would suggest that increasing the B-field
tends to decrease the DNB burning rate.  In Figure
\ref{fig:col_vs_burning_bfield}, the difference in temperature at a
given $y$ between the magnetized envelope and $B=0$ envelope is about
12\%, which gives a factor of five difference in the capture rates
between the two.  Also the transition region between the radiative
outer envelope and isothermal inner envelope and core (the sensitivity
strip) moves toward greater column or density with increasing B-field
(Ventura \& Potekhin 2001).  This would also decrease the DNB burning
rate as the burning layer moves to deeper depth and hence decreasing H
number density (at least in the case of nuclear limited DNB).  Also as
illustrated in Figure \ref{fig:EfieldDiag}, the electric field in the
$10^{12}\,{\rm G}$ magnetized envelope is on average \emph{higher}
leading up to the burning region ($\rho \sim 10^{4-5}\,{\rm
g\,cm}^{-3}$) than in the nonmagnetized envelope.  This results in a
lower number density at a given column in the magnetized envelope
($B=10^{12}\,{\rm G}$) compared to the nonmagnetized envelope, as
shown in the plot of number concentration as a function of column in
Figure \ref{fig:col_vs_burning_bfield}.  Hence, we find for a
$B=10^{12}\,{\rm G}$ field, the DNB burning rate is reduced by a
combination of a modified thermal profile and a modified electric
field, but for stronger fields the situation may be significantly
different and is a topic of further study.

\begin{figure}
\epsscale{1.0}
  \plotone{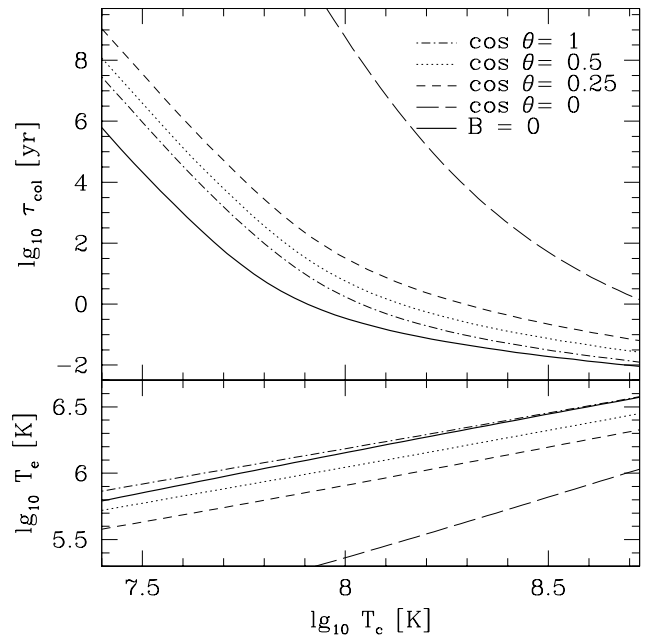}
  \caption{Hydrogen column lifetime and effective temperature for H on
    C envelopes as a function of core temperature $T_c$ for different
    magnetic field inclinations, where $\cos\theta = 1$ are the
    magnetic poles. The total H column is fixed at $y_H =
    100\,\tgsqcm$.  The magnetic field strength is $B =
    10^{12}\Gauss$, typical of young radio pulsars.  For a given core
    temperature the effective temperature of the magnetized object is
    greater for radial inclinations of the magnetic field, but the
    column lifetime of H is greater for the magnetized envelopes,
    regardless of inclination, than the nonmagnetic case (thick solid
    line).}
    \label{fig:Te_vs_tau_Bfield}
\end{figure}

We plot the H column lifetime as a function of the central core
temperature for differing inclinations of a $B=10^{12}\ {\rm G}$ with
$y_H = 100\tgsqcm$ in Figure \ref{fig:Te_vs_tau_Bfield}.  For purposes
of comparison, we also show the nonmagnetic case (thick solid line).
The low and high temperature scalings of the magnetized envelope are
similar to the scalings the nonmagnetized envelope, but the overall
normalization and transition temperature are modified by the magnetic
field.  \emph{The H column lifetime is greater for the magnetized
envelopes at $B=10^{12}\,{\rm G}$ than for the unmagnetized envelopes
for at a given $y_H$ and $T_c$, but the magnitude of the change is
highly dependent on the orientation of the magnetic field.}  In the
bottom plot, we show the effective temperature as a function of $T_c$.
The variance in column lifetime and effective temperature are mainly
due to the modification of the temperature profile by the magnetic
field especially in the conductive part of the envelope.

\begin{figure}
\epsscale{1.0} 
  \plotone{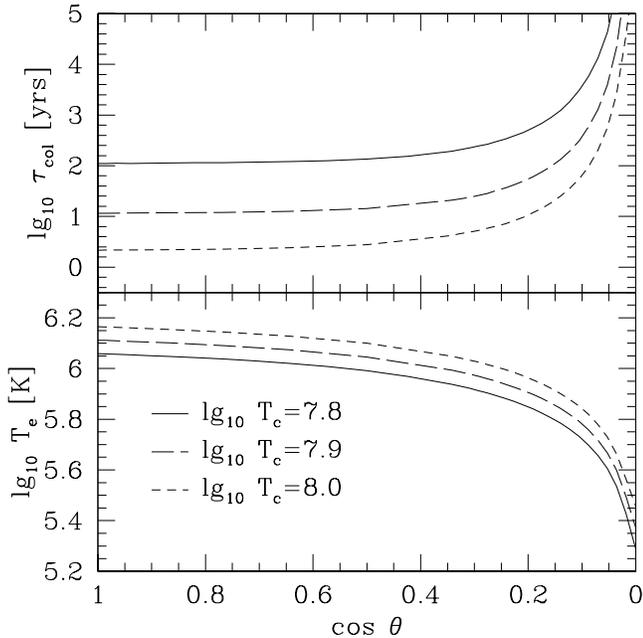}
  \caption{Hydrogen column lifetime for $y_H = 100 \tgsqcm$ and
    effective temperature for H on C envelopes as a function of
    inclination to the magnetic pole $\theta$, presuming a dipole
    field configuration, $B = B_p\left( \hat{r}\,\cos\theta +
    \hat{\theta}\,\sin\theta/2\right)$, where $B_p = 10^{12}\,{\rm
    G}$, for three different core temperatures, $T_c = 6.3, 8, 10
    \times 10^7\,\Kelvin$.  The lifetime $\tau_{\rm col}$ changes by
    almost two orders of magnitude with latitude while the effective
    temperature changes only by a factor of a few.  The variation in
    burning rate with latitude at fixed $T_c$ is due to the changing
    temperature of the burning layer due to magnetic field
    inclination. }
    \label{fig:theta_vs_tau}
\end{figure}

For a dipole geometry, Figure \ref{fig:theta_vs_tau} shows the
latitude dependence of the effective temperature and column lifetime
for fixed $T_c$, dipole field strength ($B_p = 10^{12} \Gauss$) and
total H column of $y_H = 100\tgsqcm$.  The effective
temperature and burning rate are greatest at the magnetic poles and
smallest at the equator.  At the magnetic equator the burning rate is
reduced by about two orders of magnitude, while the effective
temperature is reduced by about a factor of three.  This change in
effective temperature and column lifetime is mainly due to magnetic
field inclination changing the region where the burning layer
sits. From this plot, roughly 60\% of the NS surface has the same
burning rate (within 50\%) and effective temperature (within
30\%). Since H is preferentially depleted near the magnetic
poles, this may induce a surface current of H from the equator
to the poles, a problem that merits further study.

Much caution must be taken when looking at these results seriously
especially in the superstrong B-field case since the effects of plasma
non-ideality become more and more pronounced.  Changes in the ion and
electron equation of state due to Coulomb interactions changes the
electric field in these envelopes, possibly altering the diffusion
problem drastically.

\section{Surface Composition Evolution for Young Radio Pulsars} 

 The clear application of our theoretical work is to young radio
pulsars, which can have both the nuclear physics of DNB active, but
also, potentially mass loss via pulsars winds.  Given the uncertainty
in fallback models, the ease (in terms of availability) with which H
can be accreted or produced by fallback (i.e. spallation; Bildsten et
al. 1992), and the tiny amount of material ($\sim 10^{-19}\,{\rm
M}_{\odot}$) that comprises the NS photosphere, NS atmospheres should
comprise mainly of Hydrogen.  Furthermore, the strong gravity ($g \sim
2-3 \times 10^{14}\, {\rm cm\,s}^{-2}$) on the surface of a NS implies
that the sedimentation timescale is of order one second (Bildsten,
Chang \& Paerels 2003) and hence the photosphere should be pure H.
Hence we would expect that all young NSs possess surface H at early
times, but when the surface H begins to evolve depends on when the
fallback episode ends, which is very uncertain.

\subsection{Surface Evolution via DNB}
 
\begin{figure}
\epsscale{1.0} 
  \plotone{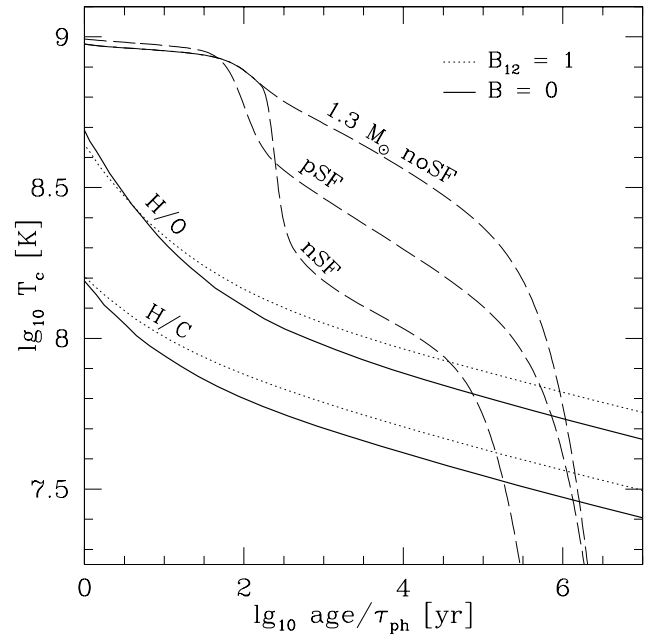}
  \caption{H column lifetime curves for H on C and H on O envelopes at
    the magnetic poles for different $B_p = 0$ (thick solid line),
    $10^{12}$ G (dotted line). Overlayed are cooling tracks
    (long-dashed lines) for various NS models (Potekhin et al. 2003)
    including standard cooling (no superfluidity, modified Urca
    cooling represented by $1.3\Msun$ noSF), proton superfluidity
    (pSF) and triplet-state core neutron superfluidity (nSF) with
    critical temperature $\approx 8 \times 10^8\,{\rm K}$.}
    \label{fig:cooling_and_DNB}
\end{figure}

 In Figure \ref{fig:cooling_and_DNB} we show the photospheric H
lifetime ($\tau_{\rm col}$ at the photosphere) curves (i.e. implied
$T_c$ for a given lifetime) for $B=0$ and $B=10^{12}\,{\rm G}$ and
proton capturing elements C and O. With the view that H could either
be initially present or added at late times, we must compare these
burning rates to the amount of time the NS spends with this $T_c$.
These times are available from NS cooling calculations, and in Figure
\ref{fig:cooling_and_DNB} we overlay a few representative cooling
tracks (long-dashed lines).  We include a standard cooling track which
presumes modified URCA cooling on a 1.3 \Msun\ NS and another with
core proton superfluidity (Potekhin et al. 2003).  D. G. Yakovlev
kindly calculated a model for us with triplet-state neutron
superfluidity in the core with a critical temperature of $8\times 10^8
\ {\rm K}$.

This figure makes clear that for H overlaying C, the burning time is
always much less than the pulsar's age for the first $10^{5-6}$
years. After this active DNB phase, any new H could stably exist on
the photosphere. For H on O, we find that the active DNB phase depends
more sensitively on the NS cooling and varies from $10^{4-6}$ yrs.
This implies that H deposited on the surface of a rapidly cooling NS
after $10^5$ years could exist stably if the underlying material is
Oxygen or heavier.  No matter what, the active DNB phase does not
naturally provide an explanation for the $10^{4-5}$ yr timescale for
surface evolution implied by the observations.

 The recent observations of mid-Z atomic lines on \oneE\ implies that
any surface Hydrogen must have been depleted on at least a
$10^{4}\,{\rm yrs}$ timescale (Sanwal et al. 2002; Mereghetti et
al. 2002; Hailey \& Mori 2002; Mori \& Hailey 2003). This pulsar is
associated with the SNR, G296.5+10; the age by association is $\approx
7\,{\rm kyrs}$ (Roger et al. 1988).  The spectral features have been
interpreted as either Oxygen lines ($B_{12}=0.55-0.75$, $z=0.06-0.21$)
or Neon lines ($B_{12}=0.8-1.1$, $z=0.62-0.86$) (Mori \& Hailey
2003). They have also been interpreted as He lines in a $B\sim
10^{14}\,{\rm G}$ field, but this would require a special magnetic
geometry in order to agree with the spindown B-field of $B_{12}
\approx 2-4$ (Pavlov et al. 2002b; Sanwal et al. 2002).\footnote{ More
recent observations indicate that there are additional lines at 2.1
and 2.8 keV (Bignami et al.  2003), which would render the
intepretation of these lines as electron cyclotron lines, but these
results require further study.  The electron cyclotron lines would
indicate a magnetic field strength of $B=6\times
10^{10}\left(1+z\right)\,{\rm G}$.  }

Let us take the interpretation of these spectral features as Oxygen
lines. Mori \& Hailey's (2003) interpretation of the spectral features
as Oxygen lines yielded $B = 5.5-7.5 \times 10^{11}\ {\rm G}$ with a
redshift of $z = 0.06-0.21$. Though it is not self-consistent, we use
the $T_e \approx 1.1 \times 10^6\,{\rm K}$ from the H atmosphere fit
of Mereghetti et al. (2002).  For Mori \& Hailey's combination of
redshift and B-field, we find that the H lifetime at the photosphere
varies from 2-20 kyrs for an underlying Oxygen layer. Hence, it is
plausible that DNB has occurred on the surface of this NS to deplete
any surface H (\emph{if initially present}) to reveal the underlying
proton-capturing element, Oxygen.

\subsection{Surface Composition Evolution via Pulsar Wind Mass Loss}

 In addition to DNB, the surface evolution of magnetized rotating NSs
may also be driven via outgoing pulsar winds. Though there is debate
about whether nuclei are present in the pulsar wind (Goldreich \&
Julian 1969; i.e. see Ruderman and Sutherland 1975), it is still
helpful to estimate how much material could possibly be excavated from
an active pulsar's surface. 
Gallant and Arons (1994) analysis of the Crab pulsar
winds showed that the ions leave that pulsar roughly at the rate of
the Goldreich-Julian current flow from the polar cap region,
\begin{equation}
\label{eq:ionrat}
\dot N_i\approx  {2\Omega^2 \mu \cos i\over ec},
\end{equation}
where $\mu$ is the NS magnetic moment.  We integrate this over time to
find the total number of excavated ions, $N=\int N_i dt$ (Michel
1975). For a constant magnetic moment (i.e. no field decay), the time
integral can be turned into an integral over the spin frequency using
the magnetic dipole spin-down law, $\dot
\Omega=-2\mu^2\Omega^3\sin^2\theta/3Ic^3$, where $I$ is the NS moment
of inertia and $\theta$ is the angle between the spin axis and
magnetic moment. This gives Michel's (1975) estimate of the total
number of ions,
\begin{equation}
\label{eq:numlos}
N_{\rm exc}\approx{3Ic^2\cos i\over \mu e \sin^2\theta}\ln\left(\Omega_i \over
\Omega_f \right),
\end{equation}
that have been excavated from the NS surface as it spins down from
$\Omega_i$ to $\Omega_f$.

Though this has never been shown, we will presume that the long
excavation timescale (of order the spin-down time) allows the polar
``hole'' left by the exiting ions to be filled in by flow of material
from other parts of the NS. Taking the moment of inertia as $I\approx
0.4 MR^2$ (Ravenhall and Pethick 1994), the excavated column depth
(presuming pure Hydrogen) is
\begin{equation}
\label{eq:yexc}
y_{\rm exc}\equiv {m_p N_i\over 4\pi R^2} \approx{Mm_pc^2\over 10\mu e}\ln\left(\Omega_i
\over
\Omega_f \right),
\end{equation}
where we dropped the angular factors. Putting in fiducial
numbers, we find a total excavated column density
\begin{equation}
\label{eq:yexcev}
y_{\rm exc}\approx 10^9 {\rm g \ cm^{-2}} \left(M\over 1.4
M_\odot\right) \left(10^{30} \ {\rm G \ cm^{3}}\over \mu\right)
\ln\left(\Omega_i \over
\Omega_f \right),
\end{equation}
which is comparable to the maximum amount of Hydrogen (Bildsten \&
Cumming 1998) or Helium (Cumming \& Bildsten 2000) that
can be present on young NSs with hot cores. 

  Though the physics of the pulsar wind is by no means well understood,
excavation has testable observational consequences for magnetized
spinning NSs.  Since $y_{\rm exc}$ is very sensitive to the NS's
magnetic dipole moment, the possible repercussions are diverse.  For
conventional radio pulsars, the underlying material may be unveiled
during the initial cooling phase ($<10^6$ yrs). For magnetars and
AXP's, a large layer of overlying material can survive on the
NS. However, for millisecond radio pulsars, the amount excavated is so
large that the rp-process ashes from the mixed Hydrogen-Helium burning
during the accretion phase (see Strohmayer \& Bildsten 2003 for an
overview) might well be revealed. 

\section{Conclusions}

 Supernova theory cannot predict the surface composition of a young
neutron star. The range of possible {\it initial} surface compositions
is large, depends on fall-back physics and residual burning at early
times (i.e. $<1-10$ years). Hence, we focused here on tracing the
abundance evolution for different initial states in the hopes of
connecting them to possible final states. This has allowed us to
construct a guide to the observer (Table \ref{table:interpretations})
on the astrophysical implications of secure detections of specific
elements. Before explaining this observational guide in detail, we
summarize our main physics results.

  We have extended our original study of DNB (CB03) to include burning
when diffusion is the rate limiting step and the important effects of
surface magnetic fields in the context of a highly approximate model
of the envelope and diffusion physics.  When diffusion is the rate
limiting step, we modified the basic equations of the envelope by
including an extra ``diffusion'' equation with an nuclear driven
sink. By including magnetic effects, we can now apply our work to
young radio pulsars.  We have shown that H can be depleted in $10^6$
years from the surface of a magnetized neutron star by diffusively
burning in deeper C layers. The timescale of this burning depends
primarily on three NS parameters: the core temperature, magnetic field
and nuclear composition of the layers where the proton captures
occur. For magnetic fields typical of young radio pulsars ($B \sim
10^{12}\,{\rm G}$) with an underlying C layer, we have shown that
the active DNB phase is relatively long (up to $10^6$ yrs). The active
DNB phase gets shorter for heavier proton capturing elements. We
applied our work to the recent observations of magnetized Oxygen lines
from \oneE\ and showed that (if initially present) Hydrogen would be
depleted from the NS surface on a timescale comparable to its age ($<
7$ kyrs), exposing a pure Oxygen layer.

 In section 4, we pointed out that the loss of ions in a pulsar wind
may be a relevant mechanism for the surface evolution of magnetized,
rotating NSs. Our estimate for the amount of material lost via the
wind is comparable to the maximum amount of Hydrogen-Helium on the
surface of the NS at birth. Though the mechanism for the wind is
highly uncertain, this amount of excavated material may have
significant observational consequences for young radio pulsars and
MSPs.

There is one remaining complication that needs to be resolved, which
is the impact of an intervening Helium layer on the H depletion
rate. The He layer would need to displace the burning layer in
order to significantly impact the rates. However, the degeneracy of
such a thick He layer allows the He to readily diffuse
downward into the underlying same $A/Z$ matter. Hence, both Coulomb
and thermal effects on the ion equation of state must be included to
resolve the miscibility of the He with the underlying layers. This
problem is also applicable to the study of stratification in
cooling white dwarfs and will be addressed separately. 
 
Given the sensitivity of DNB and excavation to the initial composition
of the envelope and physics of the pulsar wind respectively, we can
learn a tremendous amount from precision surface composition
measurements of young $B\sim 10^{12}\,{\rm G}$ neutron stars.
We have contructed a guide to understand such measurements in Table
\ref{table:interpretations}. Given the efficiency of DNB at the high
$T_c$ typical of young NSs, detection of H on a young pulsar would
indicate that either continual accretion of H rich material is
occuring, or a thick He buffer that blocks DNB is present, or there
are no proton capturing elements present underneath the H.  If H were
detected on an older pulsar ($>10^6$ yrs), this would indicate that
ion loss via pulsar wind excavation is minimal. Detection of He on a
young radio pulsar ($<10^6$ yrs) would indicate that the initial H was
depleted (via either DNB or excavation) or that the material which
fell back on the NS shortly after birth made no H.  For NS older than
$10^6$ yrs (where here we presume this is roughly the pulsar spin-down
time), detection of He would indicate that pulsar wind excavation is
ineffective at removing He from the surface.  Detection of CNO or
heavier elements may indicate that the initial H/He (and CNO for heavy
opaque material) pile was depleted via excavation (possibly DNB for
H).  Detection of CNO elements may also indicate that the outer shells
had a ``soft landing'' fallback onto the NS after birth i.e. no
spallation, if excavation is excluded.  Finally, heavy opaque
materials may also indicate lack of any fallback, if there is no
excavation.

\acknowledgements

We thank A. Y. Potekhin for clarifying discussions on the nature of
the electron EOS and opacities at high densities and for generous and
patient support throughout this work, D. G. Yakovlev for providing us
with NS cooling models, and K. Mori for providing us with atomic line
energies at large B-fields.  We thank the anonymous referee for making
very useful comments and suggestions.  We would also like to thank
P. Arras and C. J. Deloye for valuable discussions on the nature of
two-component plasmas at high densities and J. Heyl for some very
helpful suggestions and discussions.  We would also like to thank
S. W. Davis, A. Socrates, and D. Townsley for discussions. P.C. would
like to thank the Department of Physics and Department of Astronomy at
Columbia University for their hospitality. This work was supported by
the National Science Foundation under grants PHY99-07949 and
AST02-05956, NASA through grant NAG 5-8658.  L. B. is a Cottrell
Scholar of the Research Corporation.


\bibliographystyle{apj} 
\bibliography{master}

%
%

\begin{deluxetable}{c c l}
  \tablecolumns{3} 
  \tablewidth{0pt} 

  \tablecaption{Implications of Observed Surface
  Compositions of Young $10^{12}\,{\rm G}$ Neutron Stars \label{table:interpretations}}

  \tablehead{
    \colhead{Measured} & \colhead{} & \colhead{}\\
    \colhead{Surface Composition} &
    \colhead{Age of NS} & \colhead{Astrophysical Implications}}
        
  \startdata 
  
  H & $<10^6$ yrs & continual accretion\\
  & & - or - \\
  & & thick He buffer\\
  & & - or - \\
  & & no underlying p-capturing material\\
  & $> 10^6$ yrs & same as for $<10^6$ yrs\\
  & & - and possibly - \\ 
  & & no excavation from pulsar wind\\
  \hline 
  He & $<10^6$ yrs & initial H depleted via DNB/excavation\\
  & & - or - \\
  & & spallation of fallback material without H production\\
  & $>10^6$ yrs &  same as for $<10^6$ yrs\\
  & & - and possibly - \\
  & & no He excavation from pulsar wind\\
  \hline 
  C, N, O & & initial H/He depleted via DNB/excavation \\
  & & - or - \\
  & & ``soft landing'' of outer shell material (no spallation)\\
  \hline 
  Heavy opaque elements & & excavation of initial H/He/CNO pile \\
  & & - or - \\
  & & no SN fallback ($M_{\rm fallback} < 10^{-19}\Msun$)\\
  \enddata
\end{deluxetable}

\end{document}